\documentclass{article}
\usepackage [utf8] {inputenc}
\usepackage{mathtools}
\usepackage{amsthm}
\usepackage{amssymb}
\usepackage{hyperref}

\newtheorem{theorem}{Theorem}
\newtheorem{proposition}[theorem]{Proposition}
\newtheorem{lemma}[theorem]{Lemma}

\DeclareMathOperator{\X}{\mathfrak{X}}

\title{The probabilistic Weisfeiler-Leman algorithm}
\author{Saveliy V. Skresanov\footnote{The work is supported by Mathematical Center in Akademgorodok
under agreement No.~075-15-2019-1613 with the Ministry of Science and Higher Education of the Russian Federation.}}
\date{\footnotesize Sobolev Institute of Mathematics, Novosibirsk, Russia.\\ E-mail: skresan@math.nsc.ru}
\begin{document}
\maketitle

\begin{abstract}
	A probabilistic version of the Weisfeiler-Leman algorithm for
	computing the coherent closure of a colored graph is suggested.
	The algorithm is Monte Carlo and runs in time \( O(n^{1+\omega}\log^2 n) \),
	where \( n \) is the number of vertices of the graph and
	\( \omega < 2.273 \) is the matrix multiplication exponent.
\end{abstract}

\section{Introduction}

The Weisfeiler-Leman algorithm is one of the central tools in isomorphism testing.
Given a complete colored graph, the algorithm computes its coherent closure or, in other words,
the coarsest color refinement such that relations defined by the corresponding colors form
a coherent configuration.

While it was always clear that the algorithm takes polynomial time, devising an efficient implementation
was not an easy task. The best known result to this day is a \( O(n^3\log n) \)
algorithm obtained by N.~Immerman and E.~Lander~\cite{immermanWL}, see also~\cite{immermankWL};
here \( n \) is the number of vertices of the graph.

In this paper we suggest a \emph{probabilistic Weisfeiler-Leman algorithm} for computing the coherent closure,
main result being the following.

\begin{theorem}\label{mainth}
	There exists a Monte Carlo algorithm computing the coherent closure of a colored graph
	in time \( O(n^{1+\omega}\log^2 n) \), where \( n \) is the number of vertices
	and \( \omega \) is the matrix multiplication exponent.
\end{theorem}

Recall that a Monte Carlo algorithm is a randomized algorithm which may give a wrong answer
with some small prescribed probability (for example, at most \( 1/4 \)). By running a Monte
Carlo algorithm sufficient amount of times and taking a majority vote, one can make the error
probability at most \( \epsilon > 0 \) with the time cost multiplied by a factor \( O(\log(1/\epsilon)) \).

The matrix multiplication exponent \( \omega \) is less than \( 2.37286 \), see~\cite{alman},
in particular, our algorithm runs in time \( O(n^{3.37286}) \). Although this is worse than the complexity
of the best known algorithm, the algorithm we present is much easier to implement in practice. It will be seen later that
the main step of the algorithm consists essentially of multiplying two numeric matrices, which can be
done efficiently on modern hardware.

As follows from our methods, one can also obtain a probabilistic algorithm for checking if a given coloring
defines a coherent configuration or, in other words, if it is stable under the Weisfeiler-Leman algorithm.

\begin{theorem}\label{check}
	There exists a Monte Carlo algorithm checking if a given coloring of a complete graph is coherent in time \( O(n^{\omega}\log n) \),
	where \( n \) is the number of vertices and \( \omega \) is the matrix multiplication exponent.
\end{theorem}

We provide the proofs of both results in the following section. In the last section we give further remarks
and comments, including some additional estimates on time and space complexity of the algorithms, and
a note on the practical implementation of the algorithm.

\section{Proofs of main results}

We start with some conventions. In the algorithm complexity estimates we assume
the random-access machine model, where in constant time we can read and write
a memory location, perform usual arithmetic operations and generate a random
integer. Memory usage of our algorithm will be polynomial in \( n \), and
numbers involved will have bit lengths at most \( O(\log n) \), where \( n \)
is the number of vertices of the input graph.

Now, consider a complete directed graph on \( n \) vertices.
If one assigns a color to each vertex and arc of the graph, we get a \emph{coloring} \( \X \);
we will use the same symbol to denote the corresponding complete colored graph.
Observe that \( \X \) can be identified with a \( n \times n \) matrix \( X \)
where each cell contains the color of the corresponding arc (or loop, i.e.\ vertex).
If one replaces all colors by abstract (possibly noncommutative) variables \( x_1, x_2, \dots \)
we obtain a matrix \( X[x_1, x_2, \dots] \) over a polynomial ring over \( \mathbb{Z} \).
If we substitute integers \( \chi_i \in \mathbb{Z} \) in variables \( x_i \), we obtain a numeric matrix \( X[\chi_1, \chi_2, \dots] \).

Each color class of \( \X \) defines a relation on the set of vertices; we write \( S(\X) \) for the set of all such relations.
If \( A \) is an arbitrary square matrix, we can view it as a matrix of some colored graph so \( S(A) \) is also well-defined.

We say that a colored graph \( \X \) is a \emph{rainbow} if colors of its vertices
are disjoint from colors of proper arcs, and if for each relation \( s \in S(\X) \)
the transposed relation \( s^t = \{ (x, y) \mid (y, x) \in s \} \) also lies in \( S(\X) \).

Denote by \( S^\cup(\X) \) the set of all finite unions of relations from \( S(\X) \).
If a graph has two colorings \( \X \) and \( \X' \), then we say that the coloring \( \X \)
is a \emph{refinement} of \( \X' \) if \( S^\cup(\X') \subseteq S^\cup(\X) \), in other words, color classes
of \( \X \) are unions of color classes of \( \X' \).

Finally, we will always assume that there is some natural linear order on the set of colors,
so they can be identified with some interval \( \{ 1, \dots, r \} \) of integers.
\medskip

We want to compute the coherent closure of the graph \( \X \), i.e.\ the coarsest
color refinement of \( \X \) making it a coherent configuration.
Below we remind the classical Weisfeiler-Leman algorithm for computing the coherent closure~\cite{weisf};
explanations of its steps will follow shortly after.
\medskip

\noindent
\textbf{Weisfeiler-Leman algorithm}\\
\textbf{Input:} A colored graph \( \X \).\\
\textbf{Output:} The coherent closure of \( \X \).
\begin{enumerate}
	\item Refine \( \X \) to a rainbow.
	\item Let \( x_1, x_2, \dots \) be noncommutative variables.
		Compute the matrix product \( A = X[x_1, x_2, \dots]\cdot X[x_1, x_2, \dots] \)
		and refine the coloring of \( \X \) using \( A \).
	\item Repeat step~2 until the coloring stabilizes.
\end{enumerate}
\medskip

The preprocessing (step~1) can be done in the following manner. Suppose colors appearing in \( \X \) are \( \{ 1, \dots, r \} \).
If \( c(u, v) \) is the color of the arc \( (u, v) \), then set a new color \( c'(u, v) = (c(u, v), c(v, u)) \) if \( u \neq v \)
and \( c'(u, u) = (c(u, u), r+1) \) for all \( u \) and \( v \). Sort and renumber the new colors; since there
are \( n^2 \) pairs of vertices, this step can be done in \( O(n^2\log n) \) time. It is easy to see that the resulting
colored graph is a rainbow.

At the iteration step (step~2) we multiply two matrices with noncommutative variables
and a naive implementation requires \( O(n^3\log n) \) time (note that we have to sort summands
in each of the \( n^2 \) polynomials). Now, we need to refine the coloring of \( \X \)
using a matrix \( A \), meaning that we need to find the coarsest color refinement of \( \X \)
which is also a refinement of the coloring defined by \( A \).
Clearly relations of this new coloring are just (nonempty) intersections of relations from \( S(\X) \) and \( S(A) \),
so one can obtain the new coloring in the following way:
if \( c(u, v) \) is the old color of the arc \( (u, v) \), then let the new color be \( c'(u, v) = (c(u, v), A_{uv}) \).
Sorting and renumbering the new colors requires \( O(n^3 \log n) \) time since the matrix \( A \) contained polynomials
with at most \( n \) summands, so step~2 requires \( O(n^3 \log n) \) time.

Finally, the number of iterations can be bounded by the following result of M.~Lichter, I.~Ponomarenko and P.~Schweitzer:
\begin{proposition}[{\cite[Theorem~1]{lichter}}]\label{numiter}
	The classical Weisfeiler-Leman algorithm stops after \( O(n \log n) \) iterations, where \( n \) is the number of vertices
	of the input graph.
\end{proposition}
To sum up, the classical Weisfeiler-Leman algorithm complexity can be bounded by \( O(n^4\log^2 n) \),
where the slowest part is clearly the iteration step, requiring at least \( n^3 \) operations for matrix multiplication,
and \( O(n^3 \log n) \) for sorting polynomials with \( n \) summands at the recoloring phase.

The probabilistic Weisfeiler-Leman algorithm will mitigate these difficulties by replacing
noncommutative variables by independent numeric random variables, so we can apply fast matrix multiplication algorithms
and sort numbers instead of polynomials.

To state the algorithm let us first fix a universal constant \( C > 0 \), such that the classical Weisfeiler-Leman
algorithm stops after \( Cn\log n \) iterations; this constant exists by Proposition~\ref{numiter}.
The algorithm also uses an integral parameter \( m > 1 \), which will be used for estimating the probability of error.
Later we will specify how to choose this parameter to make the chance of error at most \( 1/4 \) or, in fact, as small as we want.
\medskip

\noindent
\textbf{Probabilistic Weisfeiler-Leman algorithm}\\
\textbf{Input:} A colored graph \( \X \) and a positive integer \( m \).\\
\textbf{Output:} The coherent closure of \( \X \) with the chance of error at most \( P_{err}(n, m) \).
\begin{enumerate}
	\item Refine \( \X \) to a rainbow.
	\item Choose integers \( \chi_1, \chi_2, \dots \) and \( \psi_1, \psi_2, \dots \) independently uniformly
		at random in the interval \( \{ 1, \dots, m \} \).
		Compute \( A = X[\chi_1, \chi_2, \dots]\cdot X[\psi_1, \psi_2, \dots] \)
		and refine the coloring of \( \X \) using \( A \).
	\item Make \( Cn \log n \) iterations of step~2.
\end{enumerate}
\medskip

Observe that the probabilistic Weisfeiler-Leman algorithm mimics the classical one, with the only difference being
the definition of the matrix used for refining the coloring (notice that it is a numeric matrix), and a bound on the number of iterations.
We will prove that at each step the coloring \( \X \) of the classical algorithm and the coloring \( \X' \)
of the probabilistic algorithm lead to equal color classes with high probability, i.e.\ the probability of
\( S(\X) = S(\X') \) is high. Since this is clear for the first step, it suffices to prove that
if \( A \) is the matrix arising at some step of the classical algorithm, and \( A' \) is the matrix
arising at the same step of the probabilistic algorithm, then \( A \) and \( A' \) lead to the same color classes,
i.e.\ \( S(A) = S(A') \).

Let \( x_1, x_2, \dots \) be noncommutative variables, and let \( A = X[x_1, x_2, \dots]\cdot X[x_1, x_2, \dots] \)
be the matrix arising at iteration \( k \) of the classical Weisfeiler-Leman algorithm. Let \( y_1, y_2, \dots \)
and \( z_1, z_2, \dots \) be commutative variables, and set \( B = X[y_1, y_2, \dots]\cdot X[z_1, z_2, \dots] \).
Notice that \( S(A) = S(B) \). Indeed, \( y_i \) play the role of the ``left'' variables and \( z_i \) are the ``right'' variables,
for example, a polynomial \( 2x_2x_1 + x_3x_5 \) over noncommutative variables corresponds to the polynomial
\( 2y_2z_1 + y_3z_5 \) over commutative variables.

Let \( \X \) and \( \X' \) be the colorings obtained by the classical and probabilistic Weisfeiler-Leman algorithms after \( k-1 \)
iterations; by induction, \( S(\X) = S(\X') \). Let \( A' = X'[\chi_1, \dots]\cdot X'[\psi_1, \dots] \)
be the matrix arising at iteration \( k \) of the probabilistic Weisfeiler-Leman algorithm,
and set \( B' = X'[y_1, y_2, \dots]\cdot X'[z_1, z_2, \dots] \), where \( y_i \) and \( z_i \) are abstract
commuting variables. Since \( S(\X) = S(\X') \) we obtain \( S(B) = S(B') \). We will estimate the probability
of \( S(B') = S(A') \) by the following:
\begin{lemma}[Schwartz--Zippel lemma~\cite{schwartz}]
	Let \( p \in F[x_1, x_2, \dots, x_n] \) be a non-zero polynomial of degree \( d \) over a field \( F \).
	Let \( S \) be a finite subset of \( F \). If \( \chi_1, \dots, \chi_n \in S \) are chosen at random independently
	and uniformly, then the probability of \( p(\chi_1, \dots, \chi_n) = 0 \) is at most \( d/|S| \).
\end{lemma}

Now, clearly the matrix \( A' \) can be obtained from the matrix \( B' \) by setting
variables \( y_i \) to \( \chi_i \) and \( z_i \) to \( \psi_i \), in particular, \( S^\cup(A') \subseteq S^\cup(B') \).
Suppose that this inclusion is proper, so there exist two positions in the matrix \( B' \) which contain different polynomials
\( p(y_1, \dots, z_1, \dots) \) and \( q(y_1, \dots, z_1, \dots) \) while matrix \( A' \) has equal numbers in these positions,
i.e.\ \( p(\chi_1, \dots, \psi_1, \dots) = q(\chi_1, \dots, \psi_1, \dots) \). The polynomial \( p-q \) is non-zero of degree 2,
and variables \( \chi_i, \psi_i \) are drawn from the set \( \{ 1, \dots, m \} \) independently uniformly at random,
so by the Schwartz-Zippel lemma, the probability of this event is at most \( 2/m \).
Since there are at most \( n^4 \) pairs of cells in the matrix \( B' \), the probability of \( S(A') \neq S(B') \) is at most~\( 2n^4/m \).

Now, \( S(A') = S(B') \) with probability at least \( 1 - 2n^4/m \), hence \( S(A) = S(A') \) with the same probability
by what was shown earlier. The chance that after \( k \) iterations the classical and probabilistic algorithms produce
the same color classes is at least \( (1-2n^4/m)^k \), therefore the probability of error is at most
\[ P_{err}(n, m) = 1 - \left(1-\frac{2n^4}{m}\right)^{Cn\log n}. \]
Now let us estimate this number from above.
For \( m > 2n^4 \) Bernoulli's inequality~\cite[Section~2.4]{mitrin} implies
\[ P_{err}(n, m) \leq 1 - \left(1 - \frac{2n^4}{m}Cn\log n\right) \leq \frac{2Cn^5\log n}{m}, \]
so taking \( m \geq 8Cn^5\log n \) yields \( P_{err}(m) \leq 1/4 \), as desired.

Finally, we estimate time complexity.
One can use fast matrix multiplication algorithms, hence step~2 requires time \( O(n^\omega) \), where \( \omega \) is the matrix
multiplication exponent, to compute \( A \) and time \( O(n^2\log n) \) to refine the coloring of \( \X \) using~\( A \).
Step~1 requires \( O(n^2\log n) \) time as before, and since the algorithm makes at most \( Cn\log n \)
iterations, the overall running time is at most \( O(n^{1+\omega}\log^2 n) \). 
The algorithm clearly uses polynomial amount of space, and numbers appearing in arithmetic
computations are at most \( nm^2 \) in magnitude, which amounts to \( O(\log n) \) bit length.
This concludes the proof of Theorem~\ref{mainth}.
\medskip

To prove Theorem~\ref{check}, notice that it suffices to run one iteration of the probabilistic Weisfeiler-Leman algorithm
and check if any color refinement occurred. This gives us the desired \( O(n^\omega \log n) \) time complexity.
Although we could have used the bound on the chance of error derived in the proof of Theorem~\ref{mainth},
we now show that in Theorem~\ref{check} it is possible to give a better bound, which also will be important for practical implications.

First, notice that if the input colored graph \( \X \) is a coherent configuration, the algorithm will not detect any refinement,
as it should. Suppose \( \X \) is not a coherent configuration.
Then there exist two positions in the matrix \( X \) which have equal colors, yet corresponding positions
in the matrix \( A \) have different colors (recall that \( A \) was defined in the iteration step of the classical Weisfeiler-Leman algorithm).
Reasoning as in the proof of Theorem~\ref{mainth}, corresponding positions in the matrix \( A' \) of the probabilistic algorithm
have different numbers with probability at least \( 1 - 2/m \) (notice that we need not consider all
\( n^4 \) pairs of positions). Taking \( m \geq 8 \) we can bound the chance of error by \( 1/4 \),
and the proof of Theorem~\ref{check} is complete. Observe that in this version of the argument \( m \) does not depend on \( n \) at all.

\section{Further remarks}

\noindent\textbf{Finer complexity estimates and space complexity.}
In our complexity estimates we assumed that all arithmetic operations can be performed in \( O(1) \).
Strictly speaking, this is not the case, since the value of \( m \)
is proportional to \( n^5 \) and hence bit lengths of numbers involved are \( O(\log n) \).
Therefore access to a random oracle, sum and comparison of numbers can be performed in \( O(\log n) \)
time, while multiplication can be done in \( O(\log^2 n) \) time. In this time model the complexity
of the algorithm from Theorem~\ref{mainth} can be estimated by \( O(n^{1+\omega}\log^4 n) \) (notice that
we just added a \( \log^2 n \) factor).
In Theorem~\ref{check} the value of \( m \) does not depend on \( n \), so we multiply numbers of constant
size and add numbers of size \( O(\log n) \). The complexity of the second algorithm
is thus \( O(n^\omega\log^2 n) \) in this time model.

Regarding space complexity, it can be easily seen that the classical Weisfeiler-Leman algorithm may require \( O(n^3) \) space,
since at the iteration step we may need to store a \( n \times n \) matrix of polynomials with \( n \) summands.
On the other hand, the probabilistic Weisfeiler-Leman algorithm stores only a \( n \times n \) numeric matrix,
and thus requires \( O(n^2\log n) \) space, given that the fast matrix multiplication algorithm in use requires only a quadratic amount of space.
This seems to be a mild assumption, as all Strassen-like algorithms require a quadratic amount of space.
\medskip

\noindent\textbf{Canonicity.} The classical Weisfeiler-Leman algorithm produces a \emph{canonic} color refinement
of the original colored graph \( \X \), which essentially means that colors of the resulting coherent closure depend only on the structure
and colors of the original graph, but not on the numbering of vertices. In particular, this property is useful for isomorphism testing:
if after color refinement two graphs obtained discrete colorings (i.e.\ each vertex lies in its own color class), then
the mapping sending a vertex from the first graph to the vertex from the second graph with the same color is a potential
isomorphism.

Since the probabilistic Weisfeiler-Leman algorithm uses random numbers, it does not preserve the canonicity of colors.
Nevertheless, it is still possible to use the algorithm for isomorphism testing. Given two graphs, one has to run
the probabilistic algorithm on both inputs simultaneously, using the same source of randomness. If two graphs are isomorphic,
both procedures will request the same random numbers and hence the resulting colors also will be the same.
\medskip

\noindent\textbf{Practical implementation.}
In order to implement the probabilistic Weisfeiler-Leman algorithm in practice,
it is beneficial to deviate a bit from the procedure we gave earlier. 
Since we do not know the value of the constant \( C \) from Proposition~\ref{numiter}, step~3 
of the algorithm should be replaced by
\smallskip

\noindent
\begin{center}
3'. Repeat step~2 until the coloring does not refine for \( k \) iterations.
\end{center}
\smallskip
where \( k \) is some fixed constant. As follows from the proof of Theorem~\ref{check},
if at some iteration the coloring of the graph can be refined, then the probabilistic Weisfeiler-Leman
algorithm will refine it with probability at least \( 1 - 2/m \). Therefore the probability that
\( k \) iterations will not find a proper refinement (if it exists) is at most \( (2/m)^k \). For any
\( m > 2 \) we can make this quantity as small as we like, so in the practical implementation
we can set \( m \) depending on the capacity of the numeric types used. For instance,
for a 64-bit integer the value of \( m = 10^6 \) seems appropriate; then \( k = 3 \) safety iterations
make the chance of error at most~\( 8\cdot 10^{-18} \).

Of course we cannot guarantee that after replacing step~3 with step~3' the algorithm will also terminate in \( O(n \log n) \) iterations,
so the worst-case time complexity of the practical implementation is higher.

The algorithm has been tested in practice in the computer algebra system GAP~\cite{gap},
the program will be provided to an interested reader upon request. The author has managed to
compute coherent closures of colorings on \( \simeq 1000 \) vertices in reasonable time; it seems
that this number can be considerably increased by utilizing a low-level language
and faster matrix multiplication algorithms.

\end{document}